\documentclass[12pt,a4paper,twoside]{article}
\usepackage[dvips]{graphicx}
\usepackage{color}
\usepackage{sectsty}
\sectionfont{\normalsize\bfseries\sffamily}
\subsectionfont{\normalsize\sf}
\usepackage[english]{babel}
\usepackage[margin=2.5cm,font=small,labelfont=bf]{caption}
\usepackage{geometry}
\usepackage{amsfonts,amssymb,amsthm,amsmath,mathrsfs,booktabs}
\usepackage[square]{natbib}

\newcommand{\real}{{\rm I\kern-.17em R}}

\newcommand{\esp}{\mbox{E}}

\newenvironment{keywords}                                     
    {\vspace*{3mm}                                            
    {\noindent{}\textit{Keywords\/:}}                              
        \nopagebreak\small}                                   
        {}                                                    

    {\vspace*{3mm}                                            
    {\noindent{}\textit{MSC\/:}}                              
        \nopagebreak\small}                                   
        {}                                                    

\newtheorem{theorem}{Theorem}
\newtheorem{proposition}{Proposition}                

\theoremstyle{theorem}

\theoremstyle{definition}

\def \E{\text{\rm E}}

\def \H{\text{\rm H}}

\def \cx{\le_{\rm cx}}
\def \icx{\le_{\rm icx}}

\DeclareMathOperator*{\Var}{Var}
\DeclareMathOperator*{\ZIP}{ZIP}
\begin{document}


\title{{\bf \Large Tests for zero-inflation and overdispersion\footnote{Research
supported by the Spanish MEC, grants
MTM2007-66632 and MTM2008-06281-C02-02, Comunidad de Madrid
grant S-0505/ESP/0158.\newline
E-mail addresses: amparo.baillo@uam.es, joser.berrendero@uam.es and javier.carcamo@uam.es}}}

\author{\normalsize{\sf Amparo Ba\'{i}llo, Jos{{\'{e}}} R. Berrendero and Javier C{{\'{a}}}rcamo}  \\
\normalsize{\sf Departamento de Matem{{\'{a}}}ticas, Universidad Aut{{\'{o}}}noma de Madrid,
28049 Madrid}}

\date{\normalsize\it \today}
\maketitle
\begin{abstract}
We propose a new methodology to detect zero-inflation and
overdispersion based on the comparison of the expected sample extremes
among convexly ordered distributions. The method
is very flexible and includes tests for the proportion of structural
zeros in zero-inflated models, tests to distinguish between two
ordered parametric families and a new general test to detect
overdispersion. The performance of the proposed tests is evaluated
via some simulation studies. For the well-known fetal lamb data, we conclude that  the zero-inflated Poisson model should be rejected against other more disperse models, but we cannot reject the
negative binomial model.


\end{abstract}




\newpage

\section{Introduction}

The Poisson distribution is the standard model for the analysis of
count data. However, in many situations this type of observations
exhibit a substantially larger proportion of zeros than what is
expected for the Poisson model (see Gupta
{\em et al.} (1996)). For instance, this is often the
case with count data coming from medical and public health research (see
Bohning {\it et al.} (1999) and Campbell {\textit{et al.} (1991)). 
This phenomenon usually arises when the distribution generating the data is a mixture of two
populations, the first of which yields Poisson-distributed counts
whereas the second one always contributes with a zero. 

One natural model to describe the above situation is
the so-called {\em zero-inflated Poisson} (ZIP) model. We
say that the random variable $Y(\theta,p)$ has a $\ZIP$ distribution
with parameters $\theta$ and $p$ ($\theta >0$ and $0\le p< 1$) if
\begin{equation}\label{ZIP}
\Pr\left(Y(\theta,p)=k\right)=\begin{cases}
p + (1-p)e^{-\theta/(1-p)}, & \text{if }k=0\\[0.1cm]
\displaystyle e^{-\theta/(1-p)} \frac{\theta^k}{k! (1-p)^{k-1}}, &
\text{if } k=1,2,\dots.
\end{cases}
\end{equation}
Therefore, $Y(\theta,p)$ is a mixture of a degenerate-at-zero
distribution (with weight $p$) and a Poisson distribution of mean
$\theta/(1-p)$ (with weight $1-p$). In particular,  $Y(\theta,0)$ is
the classical Poisson variable with mean $\theta$. The ZIP
distribution has been used in diverse areas such as medicine
(B{\"o}hning \textit{et al.} (1992, 1999) and van den Broek (1995)) or biology (Nie \textit{et al.} (2006)), among others. 

The expected value of the ZIP distribution is $\E
\left(Y(\theta,p)\right)=\theta$ and the variance
$\Var(Y(\theta,p))=\theta+{\theta^2 p}/{(1-p)}$ increases as $p$
increases. The zeros coming from the degenerate variable are called
\textit{structural zeros} and those from the Poisson model
\textit{sampling zeros}. It should be observed at this point that, to
keep the mean fixed for different values of $p$, we do not follow
the usual notation for the ZIP models.

If the proportion of  atypical zero observations remains undetected,
the va\-ria\-bi\-lity of the population is underestimated and the properties of standard inference
techniques are, to some extent, deteriorated. For this reason, in
the recent li\-te\-ra\-ture there are different proposals to determine
whether the Poisson model fits a data set well enough or,
alternatively, we should choose a ZIP model that allows for an extra
proportion of zero counts. A clear and concise review of several of
these tests can be found in Xie {\em et al.} (2001). A popular and
simple choice with good properties is the score test proposed by van
den Broek (1995).

Of course, as pointed out by El-Shaarawi (1985) and Thas and Rayner
(2005), the rejection of the Poisson model does not imply that the
ZIP distribution is the most appropriate model to fit the data. It
may happen that an alternative model that accounts for the observed
dispersion could fit the data better. The negative binomial and
the zero-inflated negative binomial distributions are
examples of reasonable alternatives.

In this work we introduce a new procedure to detect zero-inflation
and overdispersion. The key idea is to link the notion of
overdispersion with the concept of \textit{variability stochastic
order}. These orders arrange distributions according to their
variability (see Section 3 of Shaked and Shanthikumar (2006)).
Therefore, it is na\-tu\-ral to suppose that the observed
overdispersion is due to the data actually coming from a different
model that dominates the initially assumed distribution in a
variability order. The most important variability order is the
so-called \textit{convex order}. We use the properties of this
order to derive suitable discrepancy measures for tests in which
``overdispersion" is understood as ``convex domination".

The method we propose is flexible and easy to implement. It is based
on the empirical comparison of the expected sample extremes of two
ordered models. An important feature is that the main ideas can be
readily adapted to cover several different testing problems: tests
for the proportion of structural zeros in zero-inflated models;
procedures for testing if a parametric model is appropriate against
another one with more variability; and a new general test to detect
overdispersion. We illustrate in detail the application of the
methodology to the case of the ZIP models, but the technique can be
analogously applied in other situations.

The definitions and relevant results on stochastic convex
dominance are briefly reviewed in Section \ref{Section Convex
order}. These results supply the necessary theoretical background
for the rest of the paper. In Section \ref{Section ZIP}, we provide
a general framework to detect overdispersion in ZIP models, but
we note that the proposed method is very general and can be adapted
to many other similar scenarios. We find discrepancy measures for
tests on the proportion of structural zeros and discuss whether
the Poisson model is appropriate or we should opt for a different
model with more dispersion. In Section \ref{Section Extensions} we
establish the relationships, in terms of the convex order, for some
zero-inflated models usually considered in the literature: the
zero-inflated binomial, Poisson and negative binomial model. These
results allow to extend the previous ideas to these important
discrete models. Section \ref{Section Simulations} analyzes the
performance of the proposed tests via some Monte Carlo studies. Our
proposals are very competitive against the well-known score test in
the cases in which the latter can be applied. In Section
\ref{Section Real data}, we analyze the fetal lamb data from Leroux
and Puterman (1992) using our new procedures. For this data set we
conclude that the ZIP distribution should be rejected against other
models with more variability. This result is consistent with the
previous work by Thas and Rayner (2005). Moreover, we show that the
negative binomial model cannot be rejected. Finally, the proofs of
the main results are collected in the appendix.

\section{The convex order and overdispersion}\label{Section Convex order}

In this section, we link the overdispersion phenomenon described in
the introduction with the convex stochastic order. Given two
integrable random variables $X$ and $Y$, it is said that \textit{$X$
is less or equal to $Y$ in the convex order}, and we denote it by
$X\cx Y$, if $\E (\phi(X))\le \E (\phi(Y))$ for every convex
function $\phi$ for which the previous expectations are well
defined. Notice that, by considering the convex functions
$\phi(x)=\pm x$, the condition $X\cx Y$ implies that $\E X = \E Y$.
Furthermore, if the variables have finite second moment, applying
the definition of the convex order with $\phi(x)=(x-\E X)^2$, we
conclude that $\Var(X) \leq \Var(Y)$. Of course, establishing the
relation $X\cx Y$ is much more informative than just knowing
$\Var(X) \leq \Var(Y)$.

Roughly speaking, since convex functions take larger values when its
argument is large, if $X\cx Y$ holds, then $Y$ is more likely to
take ``extreme values" than $X$.  This idea is clear from the following
proposition. The result is a consequence of Corollary 4.A.16 and Theorem 4.A.50 in Shaked and Shanthikumar (2006),
regarding the expected value of the extreme order statistics of two ordered variables. For
$k\ge 1$, if $(X_1,\dots,X_k)$ is a random sample of size $k$ from
$X$, we denote by $X_{i:k}$ the $i$-th order statistic of the
sample, $i=1,\dots,k$. Therefore, $X_{1:k}$ and $X_{k:k}$ stand for
the minimum and maximum of the sample.

\begin{proposition}\label{Proposition expected extremes}
\label{prop:max} Let $X$ and $Y$ be integrable random variables such
that $X\cx Y$.
\begin{enumerate}
\item[\rm (a)] For all $k\ge 1$, $\E Y_{1:k}\le\E X_{1:k}$ and $\E X_{k:k}\le\E
Y_{k:k}$.

\item[\rm (b)] If for some $k\ge2$ $\E X_{1:k}= \E Y_{1:k}$ or $\E X_{k:k}= \E Y_{k:k}$, then
$X$ and $Y$ have the same distribution.

\end{enumerate}
\end{proposition}

For instance, for the ZIP variables defined as in (\ref{ZIP}), we
can prove (see Section \ref{Section Proofs}) that
\begin{equation}\label{ZIP ordering}
Y(\theta,p_1)\cx Y(\theta,p_2),\quad 0\le p_1< p_2<1,\quad
\theta>0.
\end{equation}
Hence, Proposition \ref{Proposition expected extremes} jointly with (\ref{ZIP ordering}) imply that the ZIP variable $Y(\theta,p_2)$ is expected to take \textit{strictly} larger extreme values than $Y(\theta,p_1)$ whenever $p_1<p_2$.

\section{Tests for overdispersion in ZIP models}\label{Section ZIP}

In this section we exploit Proposition \ref{Proposition expected
extremes} to derive discrepancy measures useful to test for
overdispersion in ZIP models. We emphasize that the same
technique, with the obvious modifications, can be applied in a
similar way for the zero-inflated binomial and negative binomial
models (see Section \ref{Section Extensions}) or, in general, for
any pair of ordered distributions.

The discrepancies introduced in this section are defined in terms of the em\-pi\-ri\-cal
counterparts of the expected extreme order statistics. Therefore, our goal is to detect
(significant) differences between the estimates of the expected
extremes of two distributions.

Actually, we deal with two different problems. In
Subsection~\ref{Subsection proportion} we propose statistical tests
to analyze the proportion of structural zeros in ZIP models. In
other situations, we may want to check if the ZIP model cannot
account for the dispersion of the data. Then it is adequate to apply
the nonparametric procedure of Subsection~\ref{Subsection
nonparametric}.

\subsection{Tests for the proportion of structural zeros}\label{Subsection proportion}

Given a random sample $Y_1,\dots,Y_n$ from a variable $Y(\theta,p)$
with the ZIP distribution (\ref{ZIP}), we are interested in testing
$\H_0:\, p\le p_0$ against $\H_1:\, p>p_0$, where $p_0$ is fixed and
belongs to $[0,1)$ (the left unilateral and bilateral tests may be
studied by similar arguments). There are several works in the
literature devoted to this testing problem with $p_0=0$ (see e.g.
van den Broek (1995), Xie \textit{et al.} (2001), Jansakul and Hinde
(2002) and He \textit{et al.} (2003)). This particular case is
important since it is equivalent to testing the Poisson model
against a ZIP model with a positive proportion of structural zeros.
However, as far as we know, there are no references in the
literature including tests for values of $p_0\in (0,1)$.

The method we propose is based on the following simple idea:
(\ref{ZIP ordering}) states that $Y(\theta,p_1)\cx Y(\theta,p_2)$
whenever $0\le p_1<p_2$ and hence according to Proposition
\ref{Proposition expected extremes}, the variable $Y(\theta,p)$ is
expected to take strictly larger extreme values under $\H_1$ than
under $\H_0$. Using the information in $Y_1,\dots,Y_n$, we can
estimate the expectation of the maximum (or minimum) in a generic
subsample of size $k\geq 2$ from $Y(\theta,p)$ and $Y(\theta,p_0)$.
Then, we reject $\H_0$ whenever the difference between the two
estimates is too large.

More precisely, we denote by $\E_{\theta, p}(Y_{k:k})$  and
$\E_{\theta, p}(Y_{1:k})$  the expected values of the maximum and
minimum of $k$ independent copies of $Y(\theta,p)$,
respectively.  Given the random sample $Y_1,\dots,Y_n$ from
$Y(\theta,p)$, the maximum likelihood estimates of the parameters
$\theta$ and $p$ in the ZIP model satisfy (see Johnson
\textit{et al.} 2005)
\begin{equation} \label{MLE}
\hat\theta=\bar Y = \frac{1}{n}\sum_{i=1}^n Y_i \qquad \mbox{and}
\qquad \hat p = 1 - \frac{1-n_0/n}{1-\exp(-\hat\theta/(1-\hat p))},
\end{equation}
where $n_0$ is the number of zero-counts in the sample. Then, for
$k\ge 2$, we compute the discrepancy measures:
\begin{equation}\label{discrepancy measures ZIP}
\Delta_{k:k}=\E_{\hat\theta, \hat p} (Y_{k:k})-\E_{\hat\theta, p_0} (Y_{k:k})\quad\text{ and }\quad\Delta_{1:k}=\E_{\hat\theta, p_0} (Y_{1:k})-\E_{\hat\theta, \hat p} (Y_{1:k})
\end{equation}
and reject $\H_0$ either if $\Delta_{k:k}$ or $\Delta_{1:k}$ is too
large. Observe that, from the equalities $\E (Y(\hat{\theta},p_0))=
\E (Y(\hat{\theta},\hat{p}))$ and  $\E (X_{1:2})+\E (X_{2:2})=2\,\E
(X)$ (which holds for any integrable random variable $X$), it is
readily checked that $\Delta_{2:2}=\Delta_{1:2}$.

If we denote by $F_{\theta,p}$ the distribution function of
$Y(\theta,p)$, the discrepancies in (\ref{discrepancy measures ZIP})
can be rewritten as:
\begin{eqnarray}
\Delta_{k:k} & = & \sum_{i=0}^\infty \left[
\left(F_{\hat\theta,p_0}(i)\right)^k-  \left(F_{\hat\theta,\hat
p}(i) \right)^k    \right], \nonumber \\ [-3 mm] \label{discrepancy
Poisson} \\ [-3 mm] \Delta_{1:k} & = & \sum_{i=0}^\infty \left[
\left(1-F_{\hat\theta, p_0} (i)\right)^k-  \left(1-F_{\hat\theta,
\hat p} (i)\right)^k    \right]. \nonumber
\end{eqnarray}
In practice, we can always truncate the above series to approximate their value.

To obtain the rejection region of the tests we need to find the
distribution of $\Delta_{k:k}$ or $\Delta_{1:k}$ for $k\ge 2$ under
$\H_0$. In Theorem \ref{Theorem 1} we obtain the asymptotic
distribution of $\Delta_{2:2}$ when $p_0=0$. However, in general,
the distribution of these quantities is rather involved and
a simple parametric bootstrap schema can be used instead. The
following procedure is described for the discrepancy $\Delta_{k:k}$
but the corresponding one for $\Delta_{1:k}$ is analogous:
\begin{enumerate}
\item[(a)] Find the estimate $\hat{\theta}=\bar{Y}$.
\item[(b)] Extract $B$ parametric bootstrap samples of size $n$,
$Y^*_{1,b},\dots,Y^*_{n,b}$, for $b=1,\dots,B$, from the
distribution of $Y(\hat{\theta},p_0)$.
\item[(c)] For each sample $Y^*_{1,b},\dots,Y^*_{n,b}$, obtain the estimates ${\hat{\theta}_b^*}$ and ${\hat{p}_b^*}$ using
(\ref{MLE}).
\item[(d)] Compute the discrepancies $\Delta_{k:k}^{*,b}=\E_{\hat{\theta}_b^*,\hat{p}_b^*}(Y_{k:k})-\E_{\hat{\theta}_b^*,p_0}(Y_{k:k})$, $b=1,\dots,B$.
\item[(e)] For a significance level $\alpha$, find $Q_{k:k}^{*}(\alpha)$, the $(1-\alpha)$-quantile of the values
$\{\Delta_{k:k}^{*,b},b=1,\ldots,B\}$.
\end{enumerate}
The rejection region for the test $\H_0:p\le p_0$ versus
$\H_1:p>p_0$, at significance level $\alpha$, is approximated by
\begin{equation}\label{rejection region}
R_\alpha=\{ \Delta_{k:k}> Q_{k:k}^{*}(\alpha)  \}.
\end{equation}

As it was mentioned before, the case $p_0=0$ corresponds to testing
the Poisson model against a ZIP model with $p>0$. The simulation
studies in Subsection \ref{Subsection
SimulPropStructZeros} show that $\Delta_{2:2}$ has a good behavior.
The use of $\Delta_{2:2}$ means that we compare what we expect to
obtain for the maximum (or minimum) of two independent Poisson
variables with that of two ZIP variables with $p>0$. In this case,
there is a closed-form expression for $\esp_{\theta,0}(Y_{2:2})$
(see Johnson {\em et al.} (2005), p. 166):
\begin{equation}\label{expectation of 2 Poisson}
M_2(\theta):=\esp_{\theta,0}(Y_{2:2}) = \theta + \theta
e^{-2\theta}\left(I_0(2\theta) + I_1(2\theta) \right),
\end{equation}
where $I_0$ and $I_1$ are modified Bessel functions of
the first kind (see e.g. Abramowitz and Stegun (1965)).
Using (\ref{expectation of 2 Poisson}) we can rewrite the
discrepancy $\Delta_{2:2}$ given in (\ref{discrepancy
Poisson}) with $p_0=0$ as
\begin{equation}\label{Discrepancy 2}
\Delta_{2:2} = 2 \hat p \hat \theta + (1-\hat p)^2 M_2(\hat
\theta/(1-\hat p)) - M_2(\hat \theta).
\end{equation}
This enables us to obtain the asymptotic distribution of
$\Delta_{2:2}$ under $\H_0:p=0$ (Poissonness). In the following
theorem the symbol ``$\longrightarrow_d$" stands for ``convergence
in distribution" and $N(0,1)$ is a standard normal variable.

\begin{theorem}\label{Theorem 1}
Under $\H_0:p=0$, it holds that
\[
\sqrt{n} \, \frac{\Delta_{2:2}}{\sigma(\hat \theta)}
\longrightarrow_d \mbox N(0,1), \qquad n\to \infty,
\]
where
\begin{equation}\label{standar deviation test}
\sigma^2({\hat \theta}) := \frac{\hat{\theta}^2\left(
1-e^{-2\hat\theta}\left[  (1+\hat \theta)I_0(2\hat \theta)-I_1(2\hat
\theta)+\hat \theta I_2(2\hat \theta)    \right]  \right)^2}
{e^{\hat \theta} - 1 -{\hat \theta}},
\end{equation}
and  $I_0$, $I_1$ and $I_2$ are modified Bessel functions of the first kind.
\end{theorem}
As an immediate consequence of Theorem \ref{Theorem 1}, a critical
region with asymptotic significance level $\alpha$ for $\H_0:\, p=0$
against $\H_1:\, p>0$ is
\begin{equation}\label{arr}
R_\alpha = \bigg\{\sqrt{n} \, \frac{\Delta_{2:2}}{\sigma(\hat \theta)} >
z_\alpha \bigg\},
\end{equation}
with $z_\alpha$ being the $(1-\alpha)$-quantile of the standard normal distribution.
We remark that this test is very simple and easy to implement since the Bessel functions appearing in $\Delta_{2:2}$ and $\sigma({\hat \theta})$ can be evaluated by any standard mathematical software package.

\subsection{A general test to detect overdispersion}\label{Subsection nonparametric}

Here, we deal with the problem of detecting if a data set comes from
a Poisson distribution or there is dispersion that the Poisson
model cannot take into account. The same procedure works for the
more general ZIP model or the distributions considered in Section
\ref{Section Extensions}, but we illustrate the ideas with the
Poisson distribution for the sake of simplicity.

Let us consider the family $\mathcal{P}:=\{Y(\theta) : \theta>0\}$,
where $Y(\theta)$ is a Poisson variable with mean $\theta$. We
denote by $\mathcal{P}_{\text{cx}}$ the set of all integrable random
variables, not having the Poisson distribution, that dominate in the
convex order a variable in $\mathcal{P}$. Therefore,
$\mathcal{P}_{\text{cx}}$ includes distributions with strictly more
dispersion than the Poisson variables. In particular, according to
(\ref{ZIP ordering}) and  Proposition \ref{Proposition relations 2}
in Section \ref{Section Extensions}, all the ZIP (with $p>0$) and
the (zero-inflated) negative binomial distributions are included in
$\mathcal{P}_{\text{cx}}$. Given a random sample $Y_1,\dots, Y_n$
from $Y$, we want to test $\H_0:\, Y\in \mathcal{P}$ against
$\H_1:\, Y\in \mathcal{P}_{\text{cx}}$.

In this new test the alternative hypothesis is not completely
specified in the sense that it is not given by a
parametric family. However, to handle this problem we can use
similar ideas to those in Subsection~\ref{Subsection proportion}. We
first estimate the parameter $\theta$, $\hat{\theta}=\bar{Y}$. Then,
we compute the expectation of the maximum or minimum of $k$
independent copies of $Y(\hat{\theta})$,
$\E_{\hat{\theta}}(Y_{k:k})$ and $\E_{\hat{\theta}}(Y_{1:k})$, as
before in Subsection \ref{Subsection proportion}. On the other hand,
since there is no parametric restriction under $\H_1$, we estimate $\E Y_{k:k}$ and $\E Y_{1:k}$ by means of
the following nonparametric plug-in estimators:
\begin{align*}
 \E_{F_n}(Y_{k:k}) &:=
 \sum_{i=1}^{n} \left[\left(\frac{i}{n}\right)^k-\left(\frac{i-1}{n}\right)^k\right]\, Y_{i:n},\\
  \E_{F_n}(Y_{1:k}) &:=
 \sum_{i=1}^{n} \left[\left(1-\frac{i-1}{n}\right)^k-
  \left(1-\frac{i}{n}\right)^k \right] \, Y_{i:n},
 \end{align*}
where $F_n$ is the empirical distribution function of the sample
$Y_1,\dots,Y_n$. Hence, for $k\ge 2$, we consider the discrepancies
\begin{equation}\label{Lambda discrepancies}
\Lambda_{k:k}   := \E_{F_n}(Y_{k:k})-\E_{\hat{\theta}}(Y_{k:k})
\qquad \mbox{and} \qquad \Lambda_{1:k}   :=
\E_{\hat{\theta}}(Y_{1:k})- \E_{F_n}(Y_{1:k}).
\end{equation}
Under $\H_0$ these discrepancies are close to 0 whereas, if $\H_1$
holds, then $\Lambda_{1:k}$ and $\Lambda_{k:k}$ are (strictly)
positive for $n$ large enough. Therefore, we reject $\H_0$ whenever
$\Lambda_{1:k}$ or $\Lambda_{k:k}$ are too large. The rejection
region of these tests can be derived by using a parametric bootstrap
approach similar to the one described in Subsection~\ref{Subsection proportion}.

We finally note that we actually have a different test for each
discrepancy. The power of the test may depend on the selection of the statistic. The
choice of a test with good power is addressed in
Subsection \ref{Subsection kChoice}.

\section{Extensions to other models}\label{Section Extensions}

The application of the methodology described in the previous section
relies on verifying the convex domination of the involved variables.
In this section, we establish all the relationships, according to
the convex order, among the zero-inflated versions of some commonly
used models for count data: the Poisson, the binomial
and the negative binomial models. 
For these important discrete models, these relationships allow to extend straightaway the ideas developed in the previous section.

We first note that, given a data set, it is sensible to assume that
the models that could fit the data have the same mean. Hence, all
the parametric distributions considered in this section are selected
to have the same expectation $\theta$.

For $m\ge 1$, $0\le p< 1$ and $0<\theta\le m(1-p)$, let us consider
the random variable $X(m,\theta,p)$ which is the mixture between the
degenerate-at-zero variable with weight $p$ and a binomial
variable of parameters $m$ and ${\theta}/{[m(1-p)]}$ with weight
$1-p$. In other words, $X(m,\theta,p)$ has the \textit{zero-inflated
binomial} (ZIB) distribution with probabilities
\begin{equation*}
\Pr\left(X(m,\theta,p)=k\right)=\begin{cases}
 p + \left(1-\frac{\theta}{m(1-p)}\right)^m, & \text{if }k=0\\[0.2cm]
  (1-p) {m\choose k}\left({\theta\over
m(1-p)}\right)^k\left(1-{\theta\over m(1-p)}\right)^{m-k}, &
\text{if } 1\le k\le m.
\end{cases}
\end{equation*}

Furthermore, we also consider the variable $Z(t,\theta)$ with
negative binomial (NB) distribution of parameters $1/t$ and
$t\theta$ ($t>0$ and $\theta>0$), i.e.,
\begin{equation*}
\Pr\left(Z(t,\theta)=k\right)= \displaystyle  {k+1/t-1\choose
k}\frac{(\theta t)^k}{(1+\theta t)^{k+1/t}},\quad   k\ge 0.
\end{equation*}
Among the different parametrizations of the NB distribution, we have
chosen the \textit{unique} one, $Z(t,\theta)$, with mean $\theta$
(for all $t$) and increasing in $t$ for the convex order, that is,
satisfying $Z(t_1,\theta)\cx Z(t_2,\theta)$ whenever $0<t_1<t_2$
(see Proposition \ref{Proposition relations 2} (d) below).

However, there are infinitely many possibilities to inflate with
zeros the variable $Z(t,\theta)$ preserving the mean $\theta$. Among
them, we only consider the most representative two. On the one hand,
for $t,\theta>0$ and $0\le p<1$, let $Z_1(t,\theta,p)$ be the
mixture between the degenerate-at-zero variable with weight $p$ and
the variable $Z(t(1-p),\theta/(1-p))$ with weight $1-p$. On the
other hand, for $t,\theta>0$ and $0\le p<1$ let $Z_2(t,\theta,p)$ be
the mixture between the degenerate-at-zero variable with weight $p$
and the variable $Z(t,\theta/(1-p))$ with weight $1-p$. We refer to
these two models as the \textit{zero-inflated negative binomial} (ZINB)
models.

In order to clarify the notation, Table \ref{Table models}
summarizes the relevant information about the models considered
throughout this section. We note that all the variables have a fixed
mean $\theta$ and a proportion $p$ of structural zeros.

\begin{table}[h]
\begin{center}\caption{Summary of the considered models.}\label{Table models}
\begin{tabular}{ccl}\bottomrule
\textbf{Model}  & \quad \textbf{Notation}  & \qquad \qquad
{\textbf{Variance}} \\  \toprule
  ZIB   & \quad $X(m,\theta,p)$ &
  \quad $\displaystyle \theta+\frac{\theta^2
p}{1-p}-\frac{\theta^2}{m(1-p)}$  \\  \midrule ZIP & \quad
$Y(\theta,p)$ & \quad $\displaystyle\theta+  \frac{\theta^2 p}{1-p}$
\\ \midrule
 ZINB(1) &\quad $Z_1(t,\theta,p)$  & \quad $\displaystyle\theta+ \frac{\theta^2 p}{1-p}+\theta^2 t$\\ \midrule
 ZINB(2) &\quad $Z_2(t,\theta,p)$   & \quad $\displaystyle\theta+\frac{\theta^2 p}{1-p}+\frac{\theta^2 t}{1-p}$ \\
\bottomrule
\end{tabular}
\end{center}
\end{table}

The variance of all the zero-inflated variables described before is
an increasing function of $p\in[0,1)$. Actually, the next
proposition shows that they are convexly ordered for different
values of $p$.

\begin{proposition}\label{Proposition relations 1}
Let $X(m,\theta,p)$, $Y(\theta,p)$ and $Z_i(t,\theta,p)$ ($i=1,2$)
be variables with the ZIB, ZIP and ZINB distributions
described above. If $0\le p_1< p_2<1$, then
\begin{enumerate}

\item[\rm (a)] $X(m,\theta,p_1)\cx X(m,\theta,p_2)$, for all $m\ge 1$ and $0<\theta\le m(1-p_2)$.

\item[\rm (b)] $Y(\theta,p_1)\cx Y(\theta,p_2)$, for all $\theta>0$.

\item[\rm (c)] $Z_i(t,\theta,p_1)\cx Z_i(t,\theta,p_2)$, for all $t>0$, $\theta>0$ and $i=1,2$.

\end{enumerate}

\end{proposition}

The limiting distribution of  $X(m,\theta,p)$ (as $m\uparrow\infty$) and of $Z_i(t,\theta,p)$
(as $t\downarrow 0$) for $i=1,2$ is the ZIP variable $Y(\theta,p)$. The smaller $m$ is, the more
the ZIB variable differs from the ZIP one. Also, the larger $t$ is, the more the ZINB
variables differ from the ZIP one.

For a fixed proportion of structural zeros, the next
proposition presents the relationships among these four discrete
models.
\begin{proposition}\label{Proposition relations 2}
For a fixed
$p\in[0,1)$, we have:
\begin{enumerate}

\item[\rm (a)] $X(m,\theta,p)\cx X(m+1,\theta,p)$, for all $m\ge 1$ and $0<\theta\le m(1-p)$.

\item[\rm (b)] $X(m,\theta,p)\cx Y(\theta,p)$, for all $m\ge 1$ and $0<\theta\le m(1-p)$.

\item[\rm (c)] $Y(\theta,p)\cx Z_1(t,\theta,p)\cx Z_2(t,\theta,p)$, for all $\theta>0$ and $t>0$.

\item[\rm (d)] $Z_i(t_1,\theta,p)\cx Z_i(t_2,\theta,p)$, for all $0<t_1<t_2$, $\theta>0$ and $i=1,2$.
\end{enumerate}

\end{proposition}

Proposition \ref{Proposition relations 1} allows to test on the proportion of structural zeros in all the models of this section. Further, Proposition \ref{Proposition relations 2} makes possible the comparison of these parametric families.
The nonparametric tests described in Subsection \ref{Subsection nonparametric} can also be adapted to these models.
An example of the application of these tests can be found in Section \ref{Section Real data}.

\section{Simulations}\label{Section Simulations}

We have carried out a Monte Carlo study to check the performance of
the tests described above. The simulations
also give insight into the choice of the suitable test statistic.
The significance level in all cases is fixed as $\alpha=0.05$.

\subsection{The choice of the discrepancy measure} \label{Subsection kChoice}

The approach discussed in Section \ref{Section ZIP} generates a
family of dis\-cre\-pan\-cies for the addressed testing problems. We
actually have a different test if we select the maximum or minimum
in the discrepancy: $\Delta_{k:k}$ or $\Delta_{1:k}$ in the
tests of Subsection \ref{Subsection proportion} and $\Lambda_{k:k}$
or $\Lambda_{1:k}$ in the nonparametric case of Subsection
\ref{Subsection nonparametric}. Moreover, the test statistics also differ for
each $k\ge2$. Hence, the question of finding a test with good
power arises.

Regarding the tests on the proportion of structural zeros discussed
in Subsection \ref{Subsection proportion}, observe that both
hypotheses assume that the observations follow a parametric (ZIP)
distribution. The tests mainly rely on the estimation of the
parameters of the model, and the choice of the discrepancy is of
secondary importance. Some preliminary simulations showed that
different discrepancies and values of $k$ yield similar powers.
Therefore, in this situation we opt for the simplest one
$\Delta_{2:2}=\Delta_{1:2}$ defined in (\ref{Discrepancy 2}), which
has computational advantages over the others with larger $k$'s.

We now turn to the test for overdispersion of
Subsection~\ref{Subsection nonparametric}. $\H_0$ is given by a
parametric model whereas $\H_1$ includes all the distributions that
strictly dominate an element of the initial family. Hence, $\H_1$ is
\textit{not} specified by any parametric family. In this case, the
power of the tests strongly depends both on the distribution
generating the data and on the parametric family assumed in $\H_0$.
For a fixed discrepancy, different alternatives could lead to very
different powers. Therefore, it is advantageous to have a family of
discrepancies since this provides flexibility to select a good test in each situation.


Let us briefly explain how the coefficient of variation (CV) of the
discrepancy is useful to choose a test with good properties.
Under $\H_1$, an adequate discrepancy to detect deviations from
$\H_0$ should have a large mean and low variance, that is,
a low CV. The CV of the discrepancy describes well how the
corresponding test behaves. In general, under $\H_1$, a low CV is
paralleled by a high power. This is clearly reflected in
Figure~\ref{Figure1}, where, for 1000 Monte Carlo samples, we plot
the power of the test for overdispersion for the Poisson family and
the inverse of the CV of the discrepancy $\Lambda_{1:k}$ defined in
(\ref{Lambda discrepancies}), for different values of $k$. In
Figure~\ref{Figure1}(a), the observations are generated from a ZIP
distribution $Y(3,0.05)$, while in Figure~\ref{Figure1}(b) they are
drawn from the NB distribution $Z(0.05,3)$. In the first case, a
value of $k$ around $20$ is a good choice, but in the second case
$k=2$ is clearly the best one. Therefore we use these two values of
$k$ in the simulations of Subsection \ref{Subsection Simulations
Overdispersion}.

We finally note that when analyzing only one data set, it also
becomes possible to choose a suitable discrepancy by estimating its
CV via bootstrap (see Section~\ref{Section
Real data} for details).

\subsection{Simulations for the test on the proportion of structural zeros}
\label{Subsection SimulPropStructZeros}

We consider the test on the proportion of structural zeros in a ZIP
model (Subsection~\ref{Subsection proportion}). As argued in
Subsection~\ref{Subsection kChoice}, we select $k=2$. For the
case $p_0=0$ ($\H_0$ represents the Poisson distribution), we
compare the performance of the {\em score} test (van den Broek 1995)
and the test methodology that rejects $\H_0$ if the discrepancy
$\Delta_{1:2}=\Delta_{2:2}$ in (\ref{Discrepancy 2}) is too large.
The rejection region for the latter method is chosen in two ways:
via {\em bootstrap} as in (\ref{rejection region}) and also using
the {\em asymptotic} distribution of $\Delta_{2:2}$ as in
(\ref{arr}). The number of bootstrap samples is $B=5000$.

In Table~\ref{Table1} we record the proportion of times that
$\H_0:p=0$ is rejected. For each combination of $p$ and $\theta$ in
the table, we generate 5000 Monte Carlo samples of sizes $n$ = 50,
100 and 200 from $Y(\theta,p)$. Note that our proposed procedure has a
very competitive performance in comparison to the score test. This
is specially apparent for the lowest values of $\theta$, where, when
$p>0$, in general our procedure yields a higher power than the score
test.

In Table~\ref{Table2} the results for the test
$\H_0:p\leq 0.2$ against $\H_1:p>0.2$ are displayed. In this case we only use the
procedure based on $\Delta_{2:2}$ with rejection region
(\ref{rejection region}). The number of Monte Carlo samples is again
5000.

\subsection{Simulations for the overdispersion test}
\label{Subsection Simulations Overdispersion}

We test $\H_0:\, Y\in \mathcal{P}$ ($\mathcal P$ being the Poisson
family) against $\H_1:\, Y\in \mathcal{P}_{\text{cx}}$ following the
procedure described in Subsection~\ref{Subsection nonparametric}.
The number of Monte Carlo samples is 5000 and the number of
bootstrap samples used to compute the rejection region is $B=5000$.
We generate observations with sample sizes $n$ = 50, 100 and
200, from a ZIP distribution $Y(\theta,p)$ and apply the
nonparametric procedure based on $\Lambda_{1:20}$. Afterwards, we
generate samples from the NB distribution $Z(t,\theta)$ and carry out
the test with $\Lambda_{1:2}$. Recall that the justification for selecting such discrepancies was detailed in Subsection \ref{Subsection kChoice}.
In Tables~\ref{Table3} and \ref{Table4} we display the proportion of
times that $\H_0$ is rejected. Observe how close the powers in
Table~\ref{Table3} are to those of Table~\ref{Table1}. We found this
property appealing since in this test for overdispersion no
parametric model is specified for the alternative hypothesis.


\begin{table} [h]
\caption{Proportion of times that $\H_0:p=0$ was rejected.} \label{Table1}
\begin{center}
\begin{tabular}{cccccl}
\multicolumn{2}{c}{ } & \multicolumn{3}{c}{$p$} & \\ \cline{3-5}
\multicolumn{1}{c}{$n$} & \multicolumn{1}{c}{$\theta$} &
\multicolumn{1}{c}{0} & \multicolumn{1}{c}{0.05} &
\multicolumn{1}{c}{0.1} & \\ [-1 mm] \hline\hline
    &    & 0.047 & 0.386 & 0.784 & Bootstrap \\ [-2 mm]
50  & 3  & 0.056 & 0.422 & 0.800 & Asymptotic \\ [-2 mm]
    &    & 0.036 & 0.313 & 0.722 & Score \\ [-1 mm] \hline
    &    & 0.041 & 0.768 & 0.972 & Bootstrap \\ [-2 mm]
50  & 5  & 0.055 & 0.794 & 0.981 & Asymptotic \\ [-2 mm]
    &    & 0.044 & 0.779 & 0.978 & Score \\ [-1 mm] \hline
    &    & 0.002 & 0.923 & 0.994 & Bootstrap \\ [-2 mm]
50  & 10 & 0.002 & 0.923 & 0.994 & Asymptotic \\ [-2 mm]
    &    & 0.002 & 0.923 & 0.994 & Score \\ [-1 mm] \hline
    &    & 0.052 & 0.585 & 0.964 & Bootstrap \\ [-2 mm]
100 & 3  & 0.059 & 0.604 & 0.963 & Asymptotic \\ [-2 mm]
    &    & 0.049 & 0.494 & 0.943 & Score \\ [-1 mm] \hline
    &    & 0.043 & 0.944 & 0.999 & Bootstrap \\ [-2 mm]
100 & 5  & 0.077 & 0.966 & 1.000 & Asymptotic \\ [-2 mm]
    &    & 0.045 & 0.945 & 0.999 & Score \\ [-1 mm] \hline
    &    & 0.003 & 0.994 & 1.000 & Bootstrap \\ [-2 mm]
100 & 10 & 0.003 & 0.994 & 1.000 & Asymptotic \\ [-2 mm]
    &    & 0.003 & 0.994 & 1.000 & Score \\ [-1 mm] \hline
    &    & 0.051 & 0.827 & 0.999 & Bootstrap \\ [-2 mm]
200 & 3  & 0.054 & 0.831 & 0.999 & Asymptotic \\ [-2 mm]
    &    & 0.048 & 0.762 & 0.999 & Score \\ [-1 mm] \hline
    &    & 0.050 & 0.999 & 1.000 & Bootstrap \\ [-2 mm]
200 & 5  & 0.065 & 0.999 & 1.000 & Asymptotic \\ [-2 mm]
    &    & 0.043 & 0.999 & 1.000 & Score \\ [-1 mm] \hline
    &    & 0.007 & 1.000 & 1.000 & Bootstrap \\ [-2 mm]
200 & 10 & 0.007 & 1.000 & 1.000 & Asymptotic \\ [-2 mm]
    &    & 0.007 & 1.000 & 1.000 & Score
\end{tabular}
\end{center}
\end{table}

\begin{table}[h]
\caption{Proportion of times that $\H_0:p\leq 0.2$ was rejected.} \label{Table2}
\begin{center}
\begin{tabular}{ccccc}
\multicolumn{2}{c}{ } & \multicolumn{3}{c}{$p$} \\ \cline{3-5}
\multicolumn{1}{c}{$n$} & \multicolumn{1}{c}{$\theta$} &
\multicolumn{1}{c}{0.2} & \multicolumn{1}{c}{0.25} &
\multicolumn{1}{c}{0.3} \\ [-1 mm] \hline\hline
50  & 3  & 0.069 & 0.272 & 0.581 \\ [-2 mm] 
50  & 5  & 0.067 & 0.253 & 0.554 \\ [-2 mm] 
50  & 10 & 0.061 & 0.244 & 0.546 \\ [-2 mm] 
100 & 3  & 0.062 & 0.366 & 0.781 \\ [-2 mm] 
100 & 5  & 0.065 & 0.346 & 0.780 \\ [-2 mm] 
100 & 10 & 0.061 & 0.370 & 0.774 \\ [-2 mm] 
200 & 3  & 0.061 & 0.536 & 0.953 \\ [-2 mm] 
200 & 5  & 0.065 & 0.557 & 0.962 \\ [-2 mm] 
200 & 10 & 0.069 & 0.584 & 0.963 \\
\end{tabular}
\end{center}
\end{table}

\begin{table}[h]
\caption{Proportion of rejections of $\H_0:\, Y\in \mathcal{P}$ when sampling from a
ZIP $Y(\theta,p)$.} \label{Table3}
\begin{center}
\begin{tabular}{ccccc}
\multicolumn{2}{c}{ } & \multicolumn{3}{c}{$p$} \\ \cline{3-5}
\multicolumn{1}{c}{$n$} & \multicolumn{1}{c}{$\theta$} &
\multicolumn{1}{c}{0} & \multicolumn{1}{c}{0.05} &
\multicolumn{1}{c}{0.1} \\ \hline\hline 50  & 3  & 0.043 & 0.358 &
0.780 \\ [-2 mm] 50  & 5  & 0.045 & 0.732 & 0.966 \\ [-2 mm] 50  &
10 & 0.051 & 0.901 & 0.993 \\ [-2 mm] 100 & 3  & 0.047 & 0.576 &
0.952 \\ [-2 mm] 100 & 5  & 0.052 & 0.911 & 0.999 \\ [-2 mm] 100 &
10 & 0.053 & 0.982 & 1.000 \\ [-2 mm] 200 & 3  & 0.054 & 0.839 &
0.999 \\ [-2 mm] 200 & 5  & 0.054 & 0.992 & 1.000 \\ [-2 mm] 200 &
10 & 0.050 & 0.999 & 1.000 \\ [-2 mm]
\end{tabular}
\end{center}
\end{table}

\begin{table}[h]
\caption{Proportion of rejections of $\H_0:\, Y\in \mathcal{P}$
when sampling from a NB $Z(t,\theta)$.} \label{Table4}
\begin{center}
\begin{tabular}{cccc}
\multicolumn{2}{c}{ } & \multicolumn{2}{c}{$t$} \\ \cline{3-4}
\multicolumn{1}{c}{$n$} & \multicolumn{1}{c}{$\theta$} &
\multicolumn{1}{c}{0.05} & \multicolumn{1}{c}{0.1} \\ [-1 mm]
\hline\hline 50  & 3 & 0.183 & 0.385 \\ [-2 mm] 50  & 5 & 0.309 &
0.635 \\ [-2 mm] 100 & 3 & 0.269 & 0.583 \\ [-2 mm] 100 & 5 & 0.479
& 0.871 \\ [-2 mm] 200 & 3 & 0.409 & 0.812 \\ [-2 mm] 200 & 5 &
0.710 & 0.989 \\ [-2 mm]
\end{tabular}
\end{center}
\end{table}


\begin{figure}[h]
\caption{Power (in black) of the overdispersion test for the Poisson family
and 1/CV of the discrepancy (in grey).} \label{Figure1}
\begin{center}
\resizebox{8 cm}{5
cm}{\includegraphics{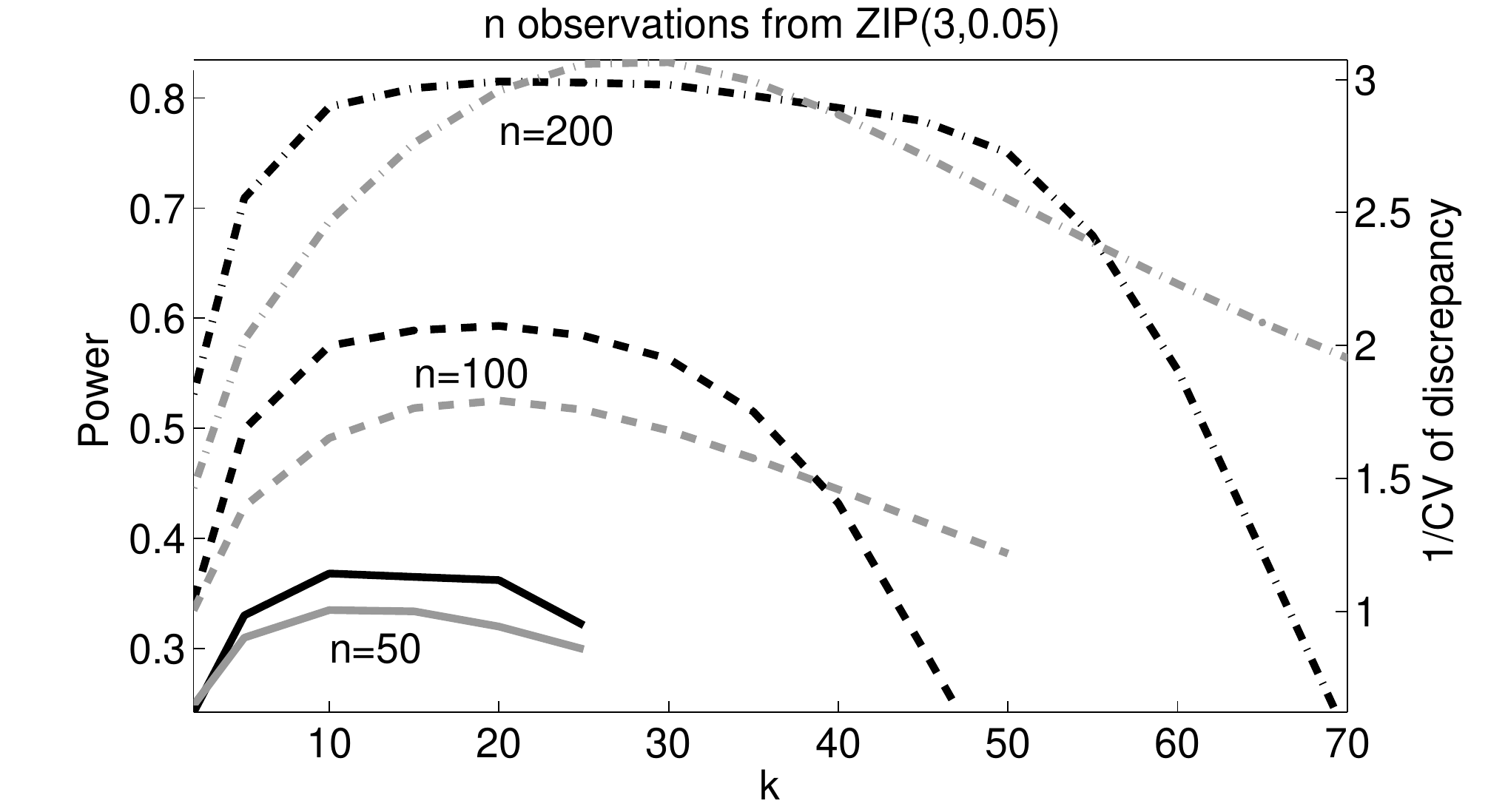}}
\\
(a) \\ [5 mm] \resizebox{8 cm}{5
cm}{\includegraphics{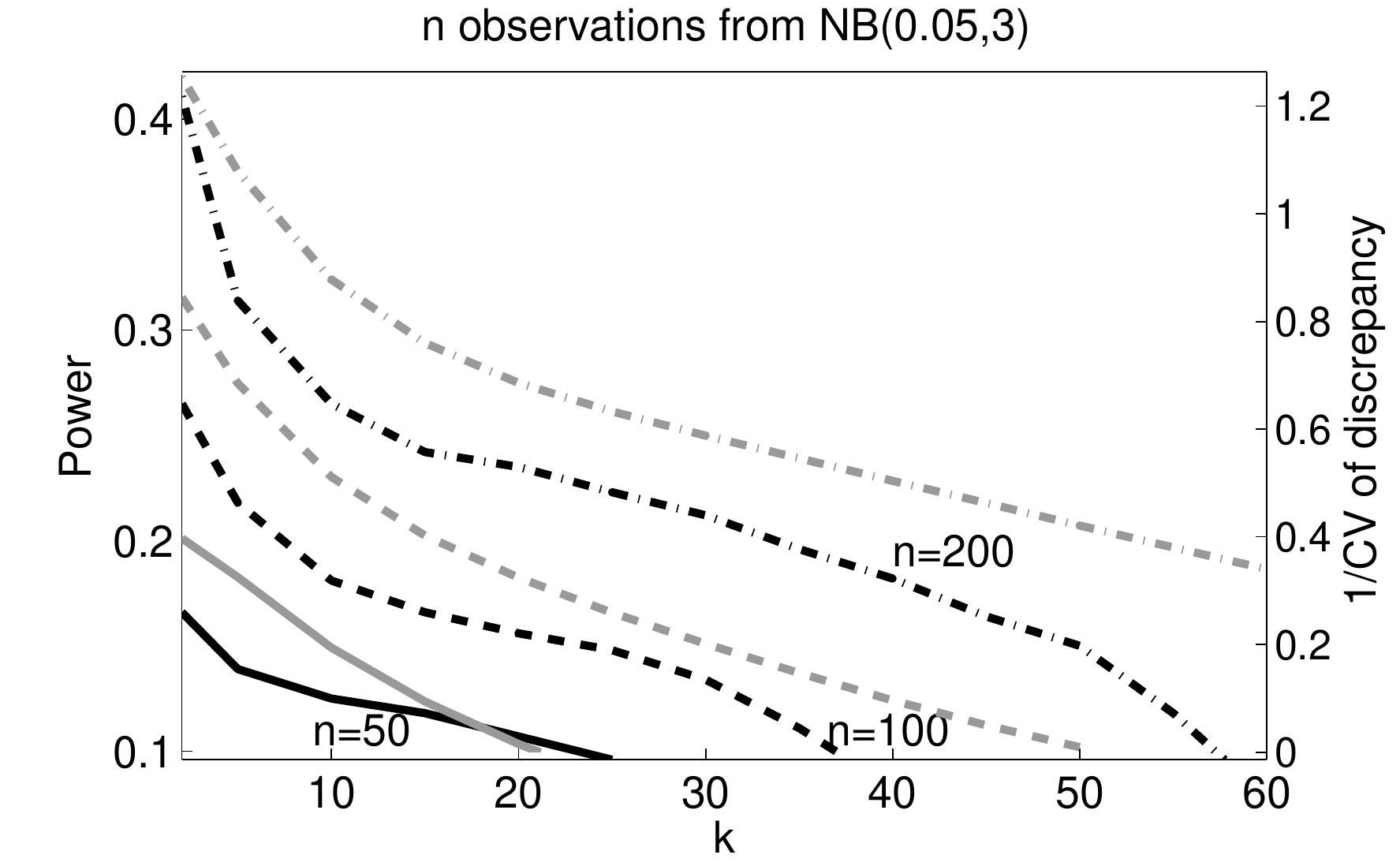}}
\\
(b)
\end{center}
\end{figure}

\section{An example with real data}\label{Section Real data}

To illustrate the usefulness of the methods proposed throughout the
paper, we analyze a data set from Leroux and Puterman (1992). The
number of movements by a fetal lamb observed through ultrasound were
recorded. We consider one particular sequence of counts of the
number of movements in each of 240 consecutive 5-second intervals
(see  Table \ref{lamb data}).

\begin{table}[h]
\begin{center}\caption{Lamb data set and expected frequencies based on Poisson,
ZIP and NB distributions.} \label{lamb data}
\begin{tabular*}{1\textwidth}{@{\extracolsep{\fill}}lcccccccc}
Outcome & 0  & 1 & 2 & 3 & 4 & 5 & 6 & 7 \\ \hline
Obs. Freq. & 182 & 41 & 12 & 2 & 2 & 0 & 0 & 1 \\  \hline
Expect. Freq. (Poisson) & 167.7 & 60.1  & 10.8 & 1.3 &  0.1 & 0.0 & 0.0 & 0.0  \\
Expect. Freq. (ZIP)     & 182.0 & 36.9  & 15.6 &  4.4  &   0.9 & 0.2 & 0.0 & 0.0 \\
Expect. Freq. (NB)      & 182.5 & 39.0  & 12.0 &  4.1 &  1.5 & 0.5 &  0.2 & 0.1
\end{tabular*}
\end{center}

\end{table}

If we assume that the data follow a Poisson distribution with mean
$\theta$, the estimate of $\theta$ is $\hat{\theta} = 0.36$. The
differences between the observed and the expected frequencies in
Table \ref{lamb data} point out that the Poisson model is
unsuitable. Douglas (1994) used the Pearson $\chi^2$ statistics to
argue that a ZIP model provides a subs\-tan\-tia\-lly improved fit. The
estimates of the parameters under the ZIP model are $\hat{\theta}=
0.36$ and $\hat{p}= 0.58$. The corresponding expected frequencies
are in the fourth row of Table~\ref{lamb data}. The fit seems indeed
better, but we could formalize this statement by testing $\H_0: p=0$
versus $\H_1: p>0$. We apply both the asymptotic
test (\ref{arr}) and the score test. Both results point out a strong
evidence (p-values below 0.0001) against the Poisson model. This
leads us to the conclusion that the ZIP distribution fits the data
much better than the Poisson one.

Rejecting the Poisson model does not necessarily imply that the ZIP
model provides the best fit. Another model could account better for
the observed dispersion. Therefore, using the nonparametric test
developed in Subsection \ref{Subsection nonparametric}, we now test
the null hypothesis that the distribution is ZIP against the
alternative that the true model has more variability than the ZIP
one. In this case, we have to select the appropriate statistics
($\Lambda_{1:k}$ or $\Lambda_{k:k}$) and a suitable value for $k$
(see Subsection \ref{Subsection kChoice}). For that purpose, we
obtain bootstrap estimates (based on 500 bootstrap samples) of the
inverse of the CV of $\Lambda_{1:k}$ and
$\Lambda_{k:k}$, for different $k$'s. The estimates as a function
of $k$ are displayed in Figure \ref{lamb cv estimates}.

\begin{figure}[h]
\begin{center}
\caption{Bootstrap estimates of $\mbox{CV}^{-1}$ for $\Lambda_{k:k}$ (solid line)
and $\Lambda_{1:k}$ (dashed line) for several values of $k$.}
\label{lamb cv estimates}
\includegraphics[height=5cm,width=8cm]{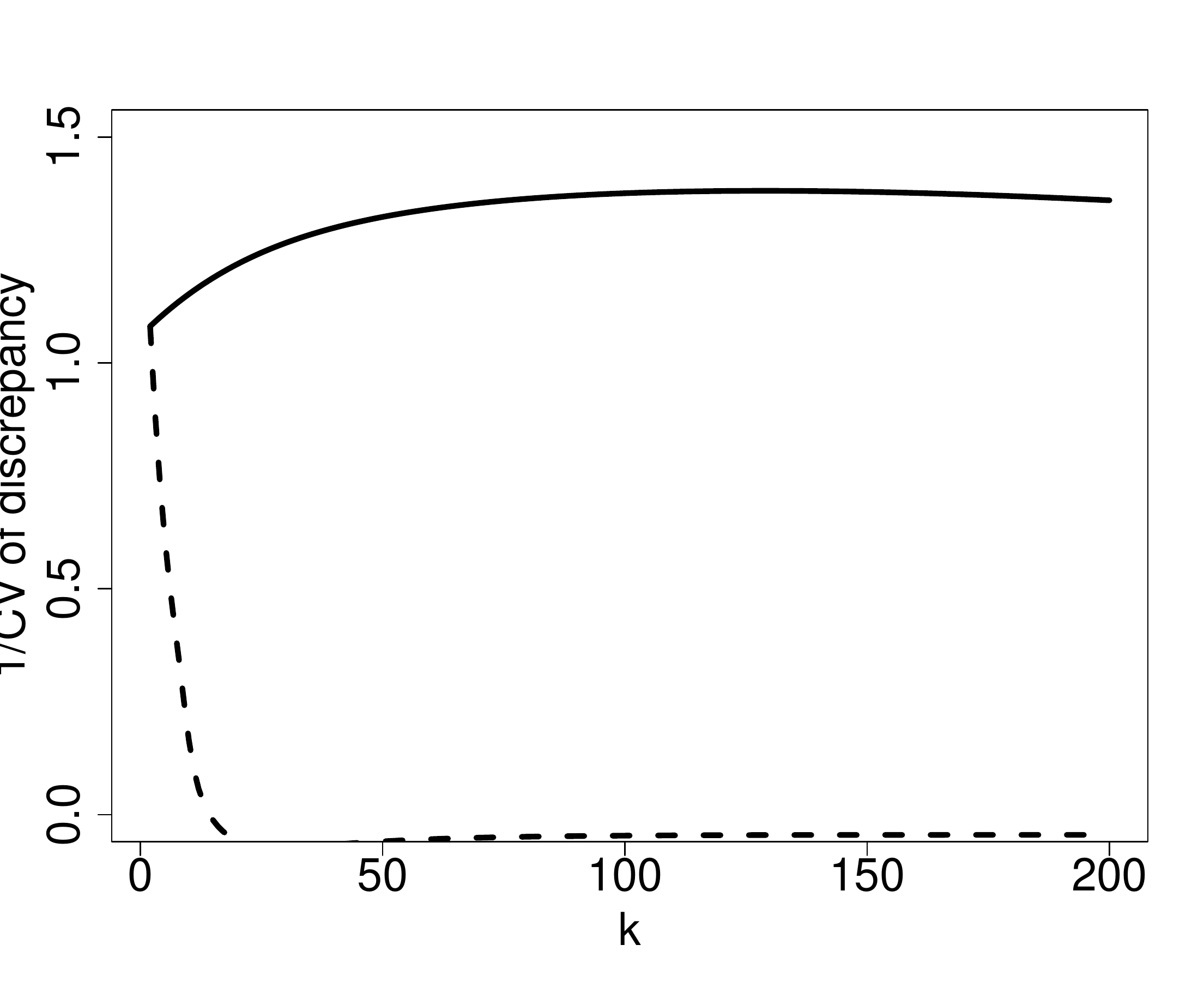}
\end{center}
\end{figure}

According to the results depicted in Figure \ref{lamb cv
estimates}, the test based on $\Lambda_{k:k}$ is pre\-fer\-able.
Moreover, for $\Lambda_{k:k}$, there is a wide
range of $k$ values (between 50 and 200, say) for which the results are fairly similar.
For the tests based on $\Lambda_{k:k}$ with $k=50,90,130$ the p-values are under 0.0005. We conclude that the ZIP model
is also clearly rejected so that other distributions with higher dispersion are more appropriate to fit this data set. Other
authors have reached the same conclusion by rather
different approaches. For instance, Ridout {\em et al.} (2001) reject
the ZIP against the ZINB using a score test in the spirit
of van den Broek (1995). Thas and Rayner (2005)
reject the ZIP against general smooth alternatives in the sense of
Neyman. A generalized Poisson distribution to fit this data set has also been
proposed by Gupta {\em et al.} (1996).

A simpler alternative to model this data is the NB distribution. The
estimated parameters are $\hat{\theta}=0.36$
and $\hat{t}=1.89$, and the corresponding expected frequencies   can
be found in the fifth row of Table~\ref{lamb data}. At first sight
it seems the fit provided by the NB is slightly better than the one
furnished by the ZIP. To confirm this feature, we adapt the
nonparametric procedure described in Subsection \ref{Subsection
nonparametric} to test the null hypothesis that the data come from a
NB distribution against the alternative that the data come from a
distribution that dominates the NB in the convex order.

We have used  bootstrap estimates of the inverse of the CV to conclude that in this case $\Lambda_{k:k}$ with $k\approx 8$ yields an appropriate test (details are omitted).
The p-values of the tests for $k= 4,6,8,10,12$ are all above 0.33.
Therefore, we cannot reject the null hypothesis and conclude that
the NB distribution accounts for the dispersion of the data better
than the ZIP model.

\section{Appendix: Proofs}\label{Section Proofs}


\subsection*{Proof of Proposition \ref{Proposition relations 1}}

We need to introduce some notation. Given two integrable random
variables $X$ and $Y$, it is said that \textit{$X$ is smaller than
$Y$ in the increasing convex order}, written $X\icx Y$, if $\E(\phi(X))\le \E(\phi(Y))$, for all increasing and convex function $\phi$, provided the expectations exist. It is easy to see that
\begin{equation}\label{cx and icx}
\text{$X\cx Y$ if and only if $X\icx Y$ and $\E X=\E
Y$.}
\end{equation}
Therefore, since all the variables considered in Proposition
\ref{Proposition relations 1} have
the same expectation $\theta$, if suffices to show that they are
ordered for the increasing convex order. Moreover, since the proof of parts (a), (b) and (c) with $i=1$ are similar, we only consider the case of ZIP variables (part (b) of the proposition).

We first note that the family $\mathcal{P}:=\{Y(\theta):\theta\in [0,\infty)\}$ , where $Y(\theta)$ is a Poisson random variable of mean $\theta\ge 0$ ($Y(0)\equiv 0$) is \textit{stochastically increasing and convex} (see Example 8.A.2 in Shaked and Shanthikumar (2006)). For $0\le p_1<p_2<1$, we define the random variables (independent of the
variables in $\mathcal{P}$) $\Theta_i=\frac{\theta}{1-p_i}B(1-p_i)$ ($i=1,2$), where
$B(1-p_i)$ is a Bernoulli variable of parameter $1-p_i$. It is readily checked that
$\Theta_1\cx \Theta_2$. Therefore, a direct application of Theorem 8.A.14 (p.
362) in Shaked and Shanthikumar (2006) yields $P(\Theta_1)\icx P(\Theta_2)$, and taking
into account (\ref{cx and icx}), we conclude
$P(\Theta_1)\cx P(\Theta_2)$. Therefore, the proof of part (b) is
finished since the ZIP variable $Y(\theta,p_i)$ has the same
distribution as $P(\Theta_i)$ ($i=1,2$).

The previous argument, based on the properties of stochastically increasing and convex families, cannot be used to prove part (c) with $i=2$ since it has not been established yet whether the collection of negative binomial variables is stochastically increasing and convex in its second parameter. We therefore need to introduce another technique inspired in the ideas used to prove Lemma 10 in de la Cal and C\'arcamo (2005). Fix $t>0$, $\theta>0$ and $0\le p_1<p_2<1$ and let $Z_2(t,\theta,p_i)$ ($i=1,2$) be the ZINB distributions defined in Section \ref{Section Extensions}. Taking into account Lemma 9 in de la Cal and  C\'arcamo (2005) and Theorem 3.A.44 (p. 133) in Shaked and Shanthikumar (2006), to prove part (c) (with $i=2$) it is enough to show that the function
\begin{equation}\label{function p}
p(k):=\Pr(Z_2(t,\theta,p_1)=k)-\Pr(Z_2(t,\theta,p_2)=k),\quad
k\ge 0,
\end{equation}
has two changes of sign, being the sign sequence $-,+,-$. To show this, we first consider the function
$$\varphi(k):=\frac{\Pr(Z_2(t,\theta,p_1)=k)}{\Pr(Z_2(t,\theta,p_2)=k)},\quad k\ge 0.$$
After some simple computations, it is easy to check that the function $f(p):=\Pr(Z_2(t,\theta,p)=0)$ is an increasing function of $p\in[0,1)$. Therefore, $\varphi(0)<1$.
Also, since
$$\frac{\varphi(k+1)}{\varphi(k)}=\frac{1-p_2+\theta t}{1-p_1+\theta t}=:c<1,\quad k\ge 1,$$
we have that $\varphi(k)=c^{k-1}\varphi(1)$ ($k\ge 1$) and this entails
$\varphi(k)\downarrow 0$ as $1\le k\uparrow\infty$. Moreover, the
equality $\sum_{k=0}^\infty
\Pr(Z_2(t,\theta,p_1)=k)=1=\sum_{k=0}^\infty \Pr(Z_2(t,\theta,p_2)=k)$
yields $\varphi(1)>1$. This implies the desired result and the proof is complete.

\subsection*{Proof of Proposition \ref{Proposition relations 2}}

In the case $p=0$, parts (a)-(d) follow from Lemmas 5 and 10 in de la Cal and C\'arcamo (2005) and Theorem 3.A.44 (p. 133) in Shaked and Shanthikumar (2006). Therefore, using that the convex order is closed under mixtures (see Theorem 3.A.12 (p. 119) of Shaked and Shanthikumar (2006)), we conclude that for any fixed $0<p<1$, (a)-(c) and the first stochastic inequality in (d) are
valid. To finish, we observe that the distribution of
$Z_1(t,\theta,p)$ is the same as the distribution of
$Z_2(t(1-p),\theta,p)$ and applying part (c) of Proposition
\ref{Proposition relations 2}, we get $Z_2(t(1-p),\theta,p)\cx
Z_2(t,\theta,p)$. This shows that $Z_1(t,\theta,p)\cx
Z_2(t,\theta,p)$ and the proof is complete.

\subsection*{Proof of Theorem 1}

We first note that the discrepancy $\Delta_{2:2}=\Delta_{2:2}(\hat \theta,\hat p)$ given in (\ref{Discrepancy 2})
is a smooth function of the maximum likelihood estimates, $\hat{\theta}$ and $\hat p$. Therefore,
the desired asymptotic distribution can be obtained
combining the  classical asymptotic theory for maximum likelihood estimators and the delta method.

According to the the asymptotic theory for maximum likelihood
estimators, we have  that:
\[
\sqrt{n} (\hat{\theta}-\theta,\hat{p}-p)^t \longrightarrow_d {N} ((0,0)^t,\Sigma),\quad n\to\infty,
\]
where ${N}((0,0)^t,\Sigma)$ is a bivariate normal distribution centered at the origin  with covariance matrix
$\Sigma$. The matrix $\Sigma$ is the inverse of the expected Fisher information
matrix, that is, $\Sigma^{-1}=-\mbox{E}_{\theta,p}[\ell''
(Y;\theta,p)]$, where $\ell''
(y;\theta,p)$ is the $2\times 2$ matrix of second partial
derivatives with respect to $\theta$ and $p$ of the log-likelihood function
$\ell(y;\theta,p)$. Using this result, after some algebra it is possible to show
that, under $\text{H}_0:\, p=0$,
\[
\sqrt{n} (\hat{\theta}-\theta,\hat{p})^t \longrightarrow_d \mbox{N} ((0,0)^t,\Sigma_0),\quad n\to\infty,
\]
where
\[
\Sigma_0 = \left(\begin{array}{cc} \theta & 0 \\
0 &  (e^\theta -1-\theta)^{-1} \end{array} \right).
\]
Now, let $\nabla \Delta_{2:2}(\theta,p)$ be the gradient of $\Delta_{2:2}(\theta,p)$
evaluated at $(\theta,p)$. Using the delta method (see e.g. van der Vaart (1998), Theorem
3.1., p. 26) we deduce that, under $\text{H}_0:\, p=0$,
\begin{equation}
\label{eq:distH}
\sqrt{n}\Delta_{2:2} \longrightarrow_d {N}(0,\sigma^2(\theta)),\quad n\to\infty,
\end{equation}
where $\sigma^2(\theta):= \nabla \Delta_{2:2}(\theta,0)^t \cdot \Sigma_0 \cdot \nabla
\Delta_{2:2}(\theta,0)$. Now we observe that
\[
\left. \frac{\partial
\Delta_{2:2}(\theta,p)}{\partial\theta}\right|_{p=0} = 0,\quad
\left. \frac{\partial \Delta_{2:2}(\theta,p)}{\partial
p}\right|_{p=0} = 2\theta - M_2(\theta) -\theta^2 e^{-2\theta}
[I_0(2\theta) - I_2(2\theta)],
\]
where the function $M_2$ is defined in (\ref{expectation of 2 Poisson}).
To obtain the last equality above we use the following properties of the modified Bessel functions
of the first kind: $I'_0(x) = I_1(x)$ and $I'_1(x) = [I_0(x) +
I_2(x)]/2$ (see Abramowitz and Stegun (1965), properties 9.6.27 and 9.6.29, p.
376). Replacing these partial derivatives and the matrix $\Sigma_0$ in the expression
$\nabla \Delta_{2:2}(\theta,0)^t \cdot \Sigma_0 \cdot \nabla
\Delta_{2:2}(\theta,0)$ yields
\begin{align*}
\sigma^2(\theta) &= \frac{(2\theta - M_2(\theta) - \theta^2e^{-2\theta}[I_0(2\theta) - I_2(2\theta)])^2}
{e^\theta - 1 -\theta}\\
&=
\frac{{\theta}^2\left(  1-e^{-2\theta}\left[  (1+ \theta)I_0(2 \theta)-I_1(2 \theta)+ \theta I_2(2 \theta)    \right]  \right)^2}
{e^{ \theta} - 1 -{ \theta}}.
\end{align*}

Finally, it is obvious that $\sigma(\hat{\theta})$ defined in (\ref{standar deviation test}) is a consistent estimator
of the standard deviation $\sigma(\theta)$. As a consequence,
from (\ref{eq:distH}) we also deduce that the conclusion of Theorem \ref{Theorem 1} holds.

\end{document}